# Shear Flow Stabilization of a z-Pinch Plasma in the Presence of a Radial Temperature Gradient

F.Winterberg

University of Nevada, Reno

**Abstract**

The previous study regarding the stabilization of a magnetized constant temperature plasma by shear flow with vorticity is extended to a plasma of non-constant temperature, where in the presence of heat source or sinks the thermomagnetic Nernst effect becomes important. Of special interest is what this effect has on the stabilization of a linear z-pinch discharge for which exact solutions are given. Solutions which are unstable for subsonic shear flow become stable if the flow is supersonic.

# 1. Introduction

The importance of thermomagnetic currents by the Nernst effect in plasma physics was early on recognized by Sakharov [1] And Braginskii [2] had pointed out that the cross-field thermoelectric force can maintain a field reversed configuration as long as a temperature gradient is sustained. Such a configuration was further studied by Ahlborn [3]. And the thermomagnetic Nernst effect for a gas embedded pinch discharge was studied by Grossmann et al. [4]. It was this paper which stimulated the author to propose the stabilization of the dense z-pinch with a thin needle-like projectile shot with supersonic velocities through the core of the pinch, which under the large magnetic pressure acting on the projectile could transform it into a dense low temperature plasma jet [5,6]. The jet along its surface generates a shear flow possessing a large vorticity. It is this vorticity containing shear flow, where the stabilizing $\rho \mathbf{v} \times \text{curl } \mathbf{v}$ force, counteracts the de-stabilizing $(c/4\pi)\ \mathbf{B} \times \text{curl } \mathbf{B}$ force. A major advancement in recognizing the importance of the thermomagnetic Nernst effect was done by Kulsrud [7], and by Hassam et al. [8]. Its importance on the shear flow of a plasma jet injected into a toroidal plasma chamber, was recognized by Hassam and Huang [9], who has noticed the similarity of their concept with the concept by the author [6].

More recently the idea to use the thermomagnetic Nernst effect for a hybrid fusion-fission reactor concept was proposed by the author [10], and was further studied by H.Motlagh [11].

Here I will show, that the previously obtained rigorous magnetohydrodynamic solutions of z-pinch configurations with superimposed axial shear flow, can be generalized to those with a radial temperature gradient with thermomagnetic currents modifying the previously obtained solutions.

## 2. The basic equations at constant temperature

In the magnetohydrodynamic approximation of a frictionless infinite conductivity plasma the equations of motion, continuity and ohm's law are

$$\frac{\partial \mathbf{v}}{\partial t} = -\frac{1}{\rho}\operatorname{grad} p - \frac{1}{2}\operatorname{grad} v^2 + \mathbf{v} \times \operatorname{curl} \mathbf{v} - \frac{1}{4\pi\rho}\mathbf{B} \times \operatorname{curl} \mathbf{B} \tag{1}$$

$$\frac{\partial \rho}{\partial t} + \operatorname{div}(\rho \mathbf{v}) = 0 \tag{2}$$

$$\frac{\partial \mathbf{B}}{\partial t} = \operatorname{curl} \mathbf{v} \times \mathbf{B} \tag{3}$$

to be supplemented by $p = p(\rho)$ and $\operatorname{div} \mathbf{B} = 0$.

Assuming cylindrical symmetry, where in a cylindrical r, φ, z coordinate system, $\partial/\partial\varphi = \partial/\partial z = 0$, exact (time independent) stationary solutions, where $\partial \mathbf{v}/\partial t = \partial \mathbf{B}/\partial t = 0$, have been found by separating the radial part of the equation of motion from its axial and azimuthal part [12]. To satisfy curl ($\mathbf{v} \times \mathbf{B}$) = 0 in the absence of an electric field, $\mathbf{v}$ must be aligned with $\mathbf{B}$. For cylindrical symmetry, $\rho = \rho(r)$, making with $v_r = 0$, $\operatorname{div}(\rho\mathbf{v}) = \rho \operatorname{div}\mathbf{v} = 0$, that is $\operatorname{div}\mathbf{v} = 0$. With $\mathbf{v}$ aligned to $\mathbf{B}$, and with $\partial/\partial\varphi = \partial/\partial z = 0$, then both $\operatorname{div}\mathbf{v} = 0$, and $\operatorname{div}\mathbf{B} = 0$.

Setting

$$\int \frac{dp}{\rho} = w = c_p T \qquad (4)$$

where $c_p$ is the specific heat at constant pressure, one can separate the equation of motion into radial part

$$\text{grad}\, w + \frac{1}{2}\text{grad}\, v^2 + \frac{B^2}{8\pi\rho^2}\text{grad}\,\rho + \mathbf{A} = 0 \qquad (5)$$

and into a second part

$$\mathbf{A} = \mathbf{v} \times \text{curl}\,\mathbf{v} - \mathbf{v}_A \times \text{curl}\,\mathbf{v}_A \qquad (6)$$

where

$$|\mathbf{A}| = A = \left(1 - \frac{1}{M_A^{\,2}(r)}\right) a(r) \qquad (7)$$

In (6) and (7)

$$\mathbf{v}_A = \frac{\mathbf{B}}{\sqrt{4\pi\rho}}$$

$$M_A = \left|\frac{\mathbf{v}}{\mathbf{v}_A}\right| \qquad (8)$$

are the Alfven velocity and Alfven Mach number.

Integration of (5) yields

$$w + \frac{1}{2}v^2 + \int \left[\frac{1}{2}v_A^2 \frac{d\ln\rho}{dr} + A(r)\right] dr = \text{const.} \tag{9}$$

With **v**×curl**v** and **v**$_A$×curl**v**$_A$ possessing only a radial component, one obtains from (6)

$$\frac{1}{2}\frac{dy}{dx} + \frac{y}{x} + \frac{1}{2}\frac{dz}{dx} = f(x) \tag{10}$$

where $x \equiv r$, $y \equiv v_\varphi^2$, $z \equiv v_z^2$, and where

$$f(x) = \frac{A(r)}{1 - \frac{1}{M_A^2(r)}} \tag{11}$$

For a given particular plasma density distribution, eq. (10) determines $v_z(r)$ in term of $v_\varphi(r)$ and vice verse, and likewise for $B_z(r)$ and $B_\varphi(r)$.

## 3. Solutions for a radial non-constant temperature distribution

In the presence of a temperature gradient, additional currents are set up in the plasma by the thermo-magnetic Nernst effect. For a fully ionized hydrogen plasma the Nernst current density is [13]

$$\mathbf{j}_N = \frac{3nkc}{2B^2} \mathbf{B} \times \nabla T \qquad (12)$$

leading to the force density

$$\mathbf{f}_N = \frac{1}{c} \mathbf{j}_N \times \mathbf{B}$$

$$= \frac{3nk}{2B^2} (\mathbf{B} \times \nabla T) \times \mathbf{B}$$

$$= \frac{3nk}{2B^2} [B^2 \nabla T - \mathbf{B}(\mathbf{B} \cdot \nabla T)] \qquad (13)$$

For a radial temperature distribution $\mathbf{B}(\mathbf{B} \cdot \nabla T) = 0$, and hence

$$\mathbf{f}_N = \frac{3}{2} nk \nabla T \qquad (14)$$

After dividing $\mathbf{f}_N$ by plasma density $\rho=nM$, the result has to be added to the r.h.s. of (5):

$$\frac{\mathbf{f}_N}{\rho} = \frac{3k\nabla T}{2M} = \frac{1}{2}c_v\nabla T = \frac{1}{2}\frac{c_p}{\gamma}\nabla T \tag{15}$$

where $\gamma=c_p/c_v=5/3$. Hence with $w=c_pT$

$$w + \frac{1}{2}\frac{c_p}{\gamma}T = 1.3c_pT = 1.3w \tag{16}$$

This simply means that in (5) gradw has to be replaced by 1.3gradw, which upon integration yields

$$1.3w + \frac{1}{2}v^2 + \int\left[\frac{1}{2}v_A{}^2\frac{d\ln\rho}{dr} + A(r)\right]dr = \text{const.} \tag{17}$$

replacing (9). Eqs. (10) and (11) remain unchanged.

## 4. The radial temperature distribution

While the effect on eq. (5), by replacing w with 1.3w appears minimal, the temperature distribution, and by implication w, can be very different in the presence of heat sources and/or sinks. A thin metallic rod placed in the center of the pinch discharge losses energy by black body radiation, and for that reason acts like a line heat sink singularity in the heat conduction equation, while the generation of thermonuclear reactions has the effect of radially distributed heat sources.

For a radial temperature distribution, the radial heat flow vector is

$$\mathbf{J} = -\kappa \frac{\partial T}{\partial r} \tag{18}$$

where $\kappa$ is the heat conduction coefficient. div$\mathbf{J} = 0$, then accounts for a line source/sink at r=0. In cylindrical coordinates one has

$$\text{div}\mathbf{J} = \frac{1}{r}\frac{\partial}{\partial r}\left(r\kappa \frac{\partial T}{\partial r}\right) = 0 \tag{19}$$

or

$$r\kappa \frac{\partial T}{\partial r} = a \tag{20}$$

where a is a constant measuring the strength of the source/sink at r=0.

For the following we assume a gyromagnetic plasma, where (ρ/B) = const. Examples are given in the Appendix. The perpendicular to the magnetic acting heat conduction coefficient $\kappa_\perp$, has been derived by Rosenbluth and Kaufman [3], and is for fully ionized hydrogen plasma given by

$$\kappa_\perp = \frac{8(\pi k)^{1/2} e^2 c^2 \ln\lambda}{3M^{3/2}} \frac{1}{T^{1/2}} \left(\frac{\rho}{B}\right)^2 \tag{21}$$

Setting

$$\kappa_\perp = ACT^{-1/2} \tag{22}$$

where

$$A = \frac{8(\pi k)^{1/2} e^2 c^2 \ln\lambda}{3M^{3/2}} \,, \qquad C = \left(\frac{\rho}{B}\right)^2 = \text{const.}$$

eq. (20) becomes

$$\frac{AC}{\sqrt{T}} \frac{dT}{dr} = \frac{a}{r} \tag{23}$$

which upon integration from $T=T_o$, $r=r_o$, to T, r, (where $r_o$ is the radius of the rod) yields

$$2AC(\sqrt{T} - \sqrt{T_o}) = a\ln\left(\frac{r}{r_o}\right) \tag{24}$$

which for $T_o=0$, corresponding to a black body radiating central metallic rod where $T_o<<T$, leads to

$$T = \left(\frac{a}{2AC}\right)^2 \left(\ln\left(\frac{r}{r_o}\right)\right)^2 \tag{25}$$

which for r→0, leads to |dT/dr|→∞.

The large temperature gradient near the surface of the rod leads there to large thermomagnetic currents. The magnetic field set up by these currents, repels the hot plasma surrounding the rod, magnetically insulating it from the hot plasma and reducing its bremsstrahlungs losses [6].

In the presence of the radial heat source h(r), one has instead of (19):

$$\frac{1}{r}\frac{\partial}{\partial r}\left(r\kappa_\perp \frac{\partial T}{\partial r}\right) = h(r) \tag{26}$$

A first integral of (26) is

$$r\kappa_\perp \frac{\partial T}{\partial r} = \int r\, h(r)\, dr \tag{27}$$

or

$$AC \frac{r}{\sqrt{T}} \frac{dT}{dr} = \int r\, h(r)\, dr \tag{28}$$

which by a second integration yields

$$2\, AC\, \sqrt{T} = \int \frac{dr}{r} \int r\, h(r)\, dr \tag{29}$$

or

$$T = \frac{1}{(2\, AC)^2} \left[ \int \frac{dr}{r} \int r\, h(r)\, dr \right]^2 + \text{const.} \tag{30}$$

## 4. Stability

Stability can most generally determined by energy principle of Bernstein, Frieman, Kruskal and Kulsrud [15]. It is most easily established for constant density ρ and if $M_A>1$. For plasmas where $β = 1$, ($β=nkT/(B^2/8π)$), $M_A>1$ means that the flow is supersonic. If ρ=const., (1) becomes

$$\frac{\partial \mathbf{v}}{\partial t} = -\nabla\left(\frac{p}{\rho}+\frac{v^2}{2}\right) + \mathbf{v}\times \text{curl}\mathbf{v} - \frac{1}{4\pi\rho}\mathbf{B}\times \text{curl}\mathbf{B} \qquad (31)$$

Or with $\mathbf{v}\|\mathbf{B}$, $v/v_A = M_A$

$$\frac{\partial \mathbf{v}}{\partial t} = -\nabla\left(\frac{p}{\rho}+\frac{v^2}{2}\right) - (1-M_A^2)\frac{1}{4\pi\rho}\mathbf{B}\times \text{curl}\mathbf{B} \qquad (32)$$

with (2) remaining unchanged.

Setting

$p \to p/\rho + v^2/2$

$\mathbf{B} \to \sqrt{1-M_A^2}\,\mathbf{B}$ \qquad (33)

and setting div v=0 for ρ=const., one obtains from the energy principle that

$$\delta W = \frac{1}{4\pi}\int (1-M_A^2)[Q^2 - (\nabla\times\mathbf{B})\cdot(\mathbf{Q}\cdot\boldsymbol{\epsilon})]d^3x > 0 \qquad (34)$$

where

$\mathbf{Q} = \nabla\times(\boldsymbol{\xi}\times\mathbf{B})$

$\boldsymbol{\xi} = \int \mathbf{v}\,dt$ \qquad (35)

Therefore, configurations which for $M_A<1$ are unstable, are stable for $M_A>1$, that means for $\beta=,1$ they are stable for supersonic flow. This conclusion is confirmed in at least one experiment done by Shumlak et al. [16, 17]. It showed that the otherwise unstable linear z-pinch, can be stabilized by a supersonic flow in the axial direction.

The condition $M_A>1$ means simply that

$$\rho v^2 > B^2/4\pi \tag{36}$$

If $B^2/4\pi \gg \rho v^2$, the magnetic pressure overwhelms the fluid stagnation pressure, and it "carriers along" the fluid, while for $\rho v^2 \gg B^2/4\pi$ the opposite is true, with the flow carrying along the magnetic field [18, 19].

If the plasma is assumed to be compressible, additional terms have to be added to the integrand of (34). They are all positive and do not lead to instabilities.

An entirely different situation arises if the flow becomes turbulent, or if through discontinuities in the plasma, either inside the plasma or by the boundary, shock waves occur, if $M_A>1$. In general, the plasma flow becomes turbulent above a critical Reynolds number. For magnetic plasma confinement turbulence is highly undesirable because it can drastically reduces the energy confinement time.

**Conclusion**

The importance of the thermomagnetic Nernst effect for the problem to stabilize a plasma by a vorticity containing shear flow, becomes important in the vicinity of cold walls where the temperature gradient is large, but can conceivably also become important in the presence of thermonuclear reactions, where the reaction rate goes with a high power of the temperature.

**Appendix**

A gyrotropic plasma is a collisionless plasma where the ion Larmor radius is small against the linear dimensions of the plasma. There the magnetic moment

$$\mu = \pi a^2 q \omega_c / 2\pi \tag{A.1}$$

is an adiabatic invariant, where $a$ is the ion Larmor radius, $q$ is the electric charge of the ion, and $\omega_c = qB/Mc$ the ion cyclotron frequency. M is the mass of the ion. With $\mu$=const. we thus have

$$a^2 B = \text{const.} \tag{A.2}$$

In the gyrotropic plasma, the particle number density is inverse proportional to $a^2$, hence

$$B/n = \text{const.} \tag{A.3}$$

This is also known as the "frozen in" condition, expressed by the Walen equation [20, 21]

$$\frac{d}{dt}\left(\frac{\mathbf{B}}{n}\right) = \left(\frac{\mathbf{B}}{n} \cdot \nabla\right) \mathbf{v} \tag{A.4}$$

For the following examples of a gyrotropic plasma we assume that the magnetic field has only a φ-component, with $|B|=B_\varphi \equiv B$, taken in a cylindrical coordinate system. There, from Maxwell's equation

$$\frac{4\pi}{c} \mathbf{j} = \text{curl} \mathbf{B} \tag{A.5}$$

we have

$$j(r) = \frac{c}{4\pi} \frac{1}{r} \frac{d}{dr}(rB) \tag{A.6}$$

and with p=2nkT, valid for a fully ionized hydrogen plasma for the magnetohydrodynamic equation

$$\nabla(2nkT) = \frac{1}{c}\mathbf{j} \times \mathbf{B} \qquad (A.7)$$

or

$$\frac{d}{dr}(nkT) = \frac{B}{8\pi}\frac{1}{r}\frac{d}{dr}(Br) \qquad (A.8)$$

for B/n = a = const., this becomes

$$\frac{d}{dr}(nkT) = \frac{an}{8\pi}\frac{1}{r}\frac{d}{dr}(Br) \qquad (A.9)$$

As a first example we assume that p = 2nkT = const., where by

$$B = \frac{const}{r} \qquad (A.10)$$

which is the magnetic field outside a current-carrying wire, valid for a magnetized plasma surrounding the wire. As a second example we assume that B=br, b=const. There then, j=cb/2π, and we find that

$$\frac{d}{dr}(2nkT) = \frac{b}{2\pi}B = \frac{b^2}{2\pi}r \qquad (A.11)$$

or

$$2nkT = \frac{b^2}{4\pi}r^2 \qquad (A.12)$$

and with n = B/a = br/a :

$$kT = \frac{ab}{8\pi}r \qquad (A.13)$$